| | |
|---|---|
| **Abstract** | Through the last decade, cold atmospheric plasma (CAP) has emerged as an innovative therapeutic option for cancer treatment. Recently, we have set up a potentially safe atmospheric pressure plasma jet device that displays antitumoral properties in a preclinical model of cholangiocarcinoma (CCA), a rare and very aggressive cancer emerging from the biliary tree with few efficient treatments. In the present study, we aimed at deciphering the molecular mechanisms underlying the antitumor effects of CAP towards CCA both in an *in vivo* and *in vitro* context. *In vivo*, using subcutaneous xenografts into immunocompromised mice, CAP treatment of CCA induced DNA lesions and tumor cell apoptosis, as evaluated by 8-oxoguanine and cleaved caspase-3 immunohistochemistry, respectively. Analysis of the tumor microenvironment showed changes in markers related to macrophage polarization. *In vitro*, incubation of CCA cells with CAP-treated culture media (i.e. plasma-activated media, PAM) led to a dose response decrease in cell survival. At molecular level, CAP treatment induced double-strand DNA breaks, followed by an increased phosphorylation and activation of the cell cycle master regulators CHK1 and p53, leading to cell cycle arrest and cell death by apoptosis. In conclusion, CAP is a novel therapeutic option to consider for CCA in the future. |
| **Keywords** | Cholangiocarcinoma; cold plasma; innovative therapy; tumor cells; macrophages; plasma selectivity; plasma jet |

# I. Introduction

Cholangiocarcinoma (CCA) is a tumor of the biliary tree with poor prognosis characterized by a dense desmoplastic stroma [1]. CCA is a rare tumor. Currently CCA accounts for 3% of all gastrointestinal cancers but overall, its incidence tends to increase worldwide. So far, surgical resection of the tumor is the only curative and effective therapeutic option. However, this cancer is usually diagnosed at advanced stage so that this treatment is feasible in a small proportion of patients and recurrence is high. When tumor resection is not possible or when recurrence occurs, the therapeutic alternatives consist in palliative treatments based on chemotherapy regimens with poor results [2]. Hence, there is a need of new therapeutic approaches.

Cold atmospheric plasma (CAP) (named also non-thermal plasma or low temperature plasma) is a weakly ionized gas created by electrical discharges, composed of transient, energetic and chemical active species (electrons, ions, metastables, radicals) that displays radiation, gas dynamics and electric field properties. Today, CAP interaction with biological systems (cells, tissues, tumors) is studied to address medical issues such as blood clotting, wound healing, dentistry, repair surgery, cosmetics, infectious and inflammatory diseases, and oncology [3]. CAP science and technology appear as a new research avenue to provide breakthrough solutions where conventional therapies in cancer appear limited [3]. Indeed, plasmas can reduce cell proliferation or tumor volume in preclinical mice models, in several types of cancers including skin, pancreatic, bladder, colon [4,5]. Therefore, plasmas have major potential in driving antitumor effects, notably in resistant tumors such as CCA. The primary action of CAP is to generate long-lived molecules such as reactive oxygen and nitrogen species (RONS) mainly from nitrogen and oxygen in atmospheric air or solution. This action can be either beneficial or deleterious on living tissues depending on their concentrations. RONS are primarily responsible for anti-tumor activity of CAP. They drive cell cycle arrest and cell death by damaging DNA and regulating cancer-relevant molecules such as the tumor suppressor p53 [6,7].

To date, only two studies addressed the potential of CAP to treat liver tumors [5,8]. In these studies, CAP was tested on hepatocellular carcinoma cell lines and induced cell death. To investigate CAP as a potential new therapeutic option, we previously engineered a new cold plasma jet device that showed significant antitumor effects in a mouse CCA model, without inducing toxic effects on heathy tissue [9]. Here, we aim to gain insight into the molecular mechanisms by which CAP halts CCA development and progression *in vivo* and *in vitro*. In addition, we investigated whether CAP has an effect on non-tumoral cells notably hepatocytes, the parenchymal liver cells. Evidence was previously provided to indicate that CAP induced cell death selectively in tumor cells and not in non-malignant cells [3]. The tumor itself is a complex tissue structure including cells of the tumor microenvironment such as cancer-associated fibroblasts (CAF), endothelial cells (EC) and tumor-associated macrophages (TAM). Therefore, we also evaluated *in vivo* the impact of CAP on these cell populations.

## II. Results

### II.1. CAP treatment reduces cholangiocarcinoma progression in a murine xenograft model

We previously compared two CAP generating devices, *i.e.* Plasma Gun (PG) and Plasma Tesla Jet (PTJ), showing that both devices were safe but differed with respect to anticancer properties [9]. Only PTJ (Figure 1a) displayed a significant therapeutic efficacy in a subcutaneous xenograft model of CCA [9]. In the present study, we used the same model to further analyze the molecular mechanisms accounting for PTJ effects in the same preclinical model. To better assess the effect of CAP on CCA growth, we compared its effect with that of gemcitabine, one of the chemotherapeutic drugs currently used in CCA patient treatment. EGI-1 CCA cells were injected to induce tumors in the flank of immunodeficient mice and once the tumors reached an arbitrary volume of 200 mm$^3$, we applied CAP directly on the tumors (Figure 1b) or we administrated gemcitabine by intraperitoneal injection twice a week for 3 weeks (see red arrows in Figure 1c). Animals were sacrificed 2 hours after the last treatment. As shown in Figure 3 c-e, tumor size and growth rate were significantly reduced after the application of CAP consistently with our previous results [9]. The well-established antitumoral effect of gemcitabine was evident and exceed that of CAP [10]. To verify that local CAP treatment did not induce side effects in the whole organism, we measured the plasma concentrations of alanine aminotransferase (ALAT) and aspartate aminotransferase (ASAT) as well as lactate dehydrogenase (LDH) in treated mice. No significant difference of concentration was observed between CAP treated animals and controls. By contrast, ASAT and LDH were significantly increased in the animals that received gemcitabine, indicating liver damage. These results show the advantage of direct CAP treatment which remains local over the systemic effects of gemcitabine but also less toxic. If, at first sight, CAP may appear less efficient than gemcitabine, one has to underline that CAP exposure times were as low as 1 min while the lifetime of gemcitabine injected in the organism is several hours.

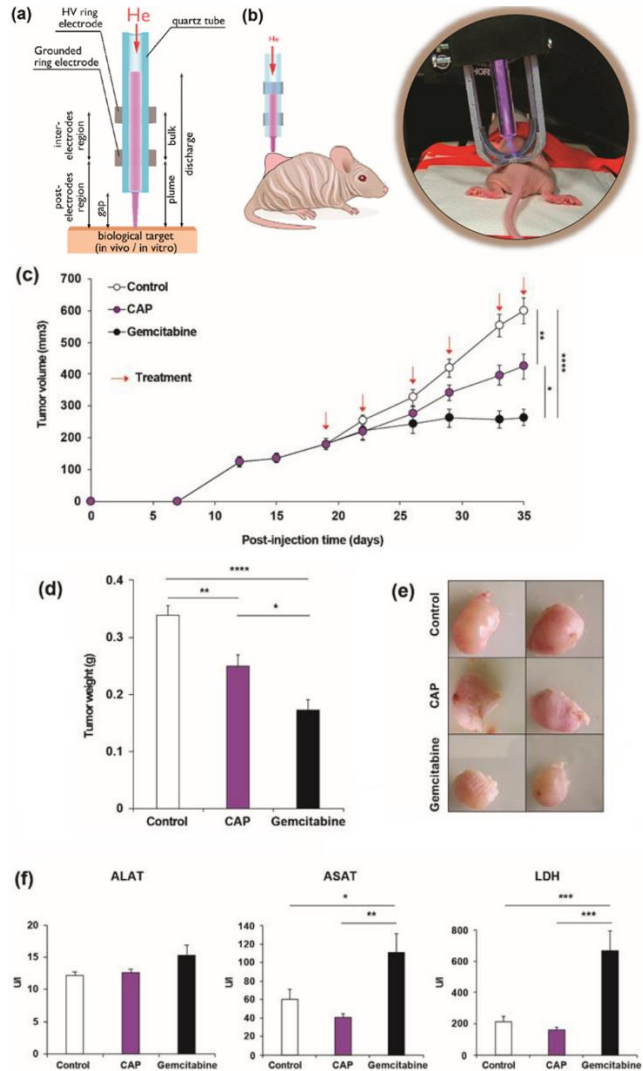

*Figure 1. (a) Experimental setup of the Plasma Tesla Jet device (PTJ). (b) Schematic representation and representative image of the cold atmospheric plasma (CAP) application to subcutaneous xenograft CCA tumors. (c) Tumor volume of mice bearing CCA developed from EGI-1 cells treated with gemcitabine (120 mg/kg, black circles), CAP (1 min at 9kV of amplitude, frequency=30kHz, duty cycle=14%, gap=10 mm, purple circles) or untreated (control, white circles). Arrows indicate treatments points with CAP and gemcitabine. (d) Tumor weight at sacrifice (day 35). (e) Representative images of tumors from each group at sacrifice. (f) Plasmatic concentrations of alanine aminotransferase (ALAT), aspartate aminotransferase (ASAT) and lactate dehydrogenase (LDH). Values are expressed as means ± SEM. \*, p < 0.05; \*\*, p < 0.01; \*\*\*, p < 0.001; \*\*\*\*, p < 0.0001.*

## II.2. Cold atmospheric plasma induces apoptosis in cholangiocarcinoma cells in vivo

To further evaluate the effect of CAP on CCA xenografts, we performed a histological analysis of the tumors. A deep analysis revealed the presence of purple round structures that represent calcifications (Figure 2a and b). These calcifications are often associated with apoptotic bodies and may represent a late state of condensed apoptotic structures. The quantification showed an increased number of calcifications in tumors treated with CAP or gemcitabine compared to controls (Figure 2c).

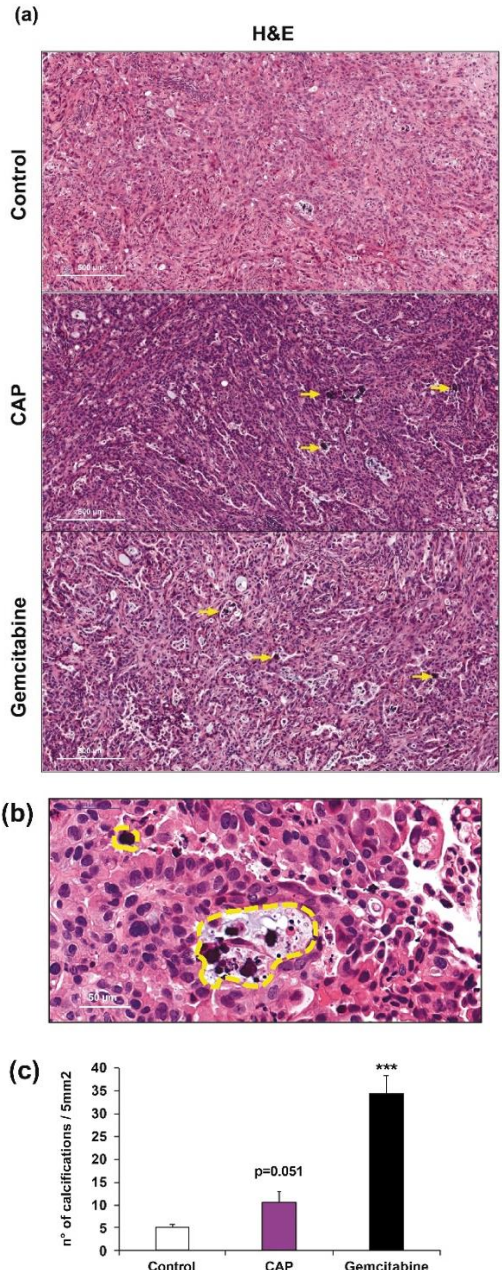

*Figure 2. (a) Representative HE staining of control (upper panel), CAP (middle panel) and gemcitabine (bottom panel) treated xenograft tumors. Magnification x125. Scale: 500 µm. (b) Magnification (x1000) of calcifications corresponding to apoptotic bodies (outlined in yellow). Scale: 50 µm. (c) Quantification of apoptotic structures. \*\*\*, p < 0.001; compared with control tumors.*

The presence of these calcifications prompted us to study apoptosis, the main type of cell death related to CAP, by performing immunostaining against cleaved caspase-3 (cCaspase-3), a critical executioner of apoptosis that is responsible for the cleavage of many key proteins. As shown in Figure 3 (left panels), animals treated with CAP showed an intense staining of cCaspase-3 in some areas of the tumors compared to the controls. This staining was also present but weaker in animals that received gemcitabine and even stronger in the CAP group. These differences that can be explained by the time at which the animals were sacrificed, i.e. approximately 2 hours after their CAP or gemcitabine treatments. Since CAP is applied locally, its effects operate faster than drugs delivered intraperitoneally such as gemcitabine. Indeed, this drug must be first absorbed and then transported to the tumors. In that latter case, the therapeutic effects of gemcitabine may be observed later than 2 hours.

Since one of the main effects of CAP is the production of RONS, we evaluated the presence of cellular components altered as a result of reactive species overload, more specifically 8-oxoguanine, one of the major products of DNA oxidation, as an event that could unchain the signaling pathways leading to cell death by apoptosis. As shown in Figure 3 (right panels), CAP treatment was able to strongly induce DNA alterations. In addition, it is worth noting that these alterations were colocalized with the areas positive for cleaved caspase-3 (left panels). This perfect overlapping enables us to bridge DNA damage with cell apoptosis. Interestingly, there was no staining of 8-oxoguanine in tumors from the group that received gemcitabine, showing that the main effects of this drug are not mediated by reactive species related molecular mechanisms.

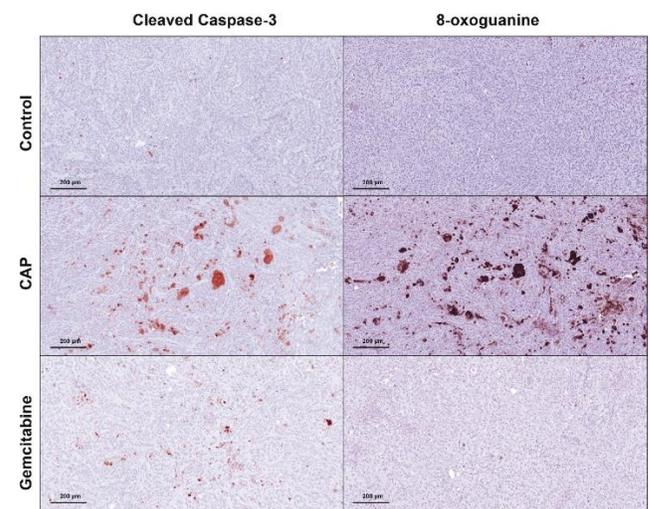

*Figure 3. Representative IHC staining of cleaved caspase-3 and 8-oxoguanine in control (upper panel), CAP (middle panel) and gemcitabine (bottom panel) treated xenograft tumors. Magnification, x250. Scale: 200 µm.*

## II.3. Cold atmospheric plasma reduces viability of cholangiocarcinoma cells but not of normal hepatocytes *in vitro*

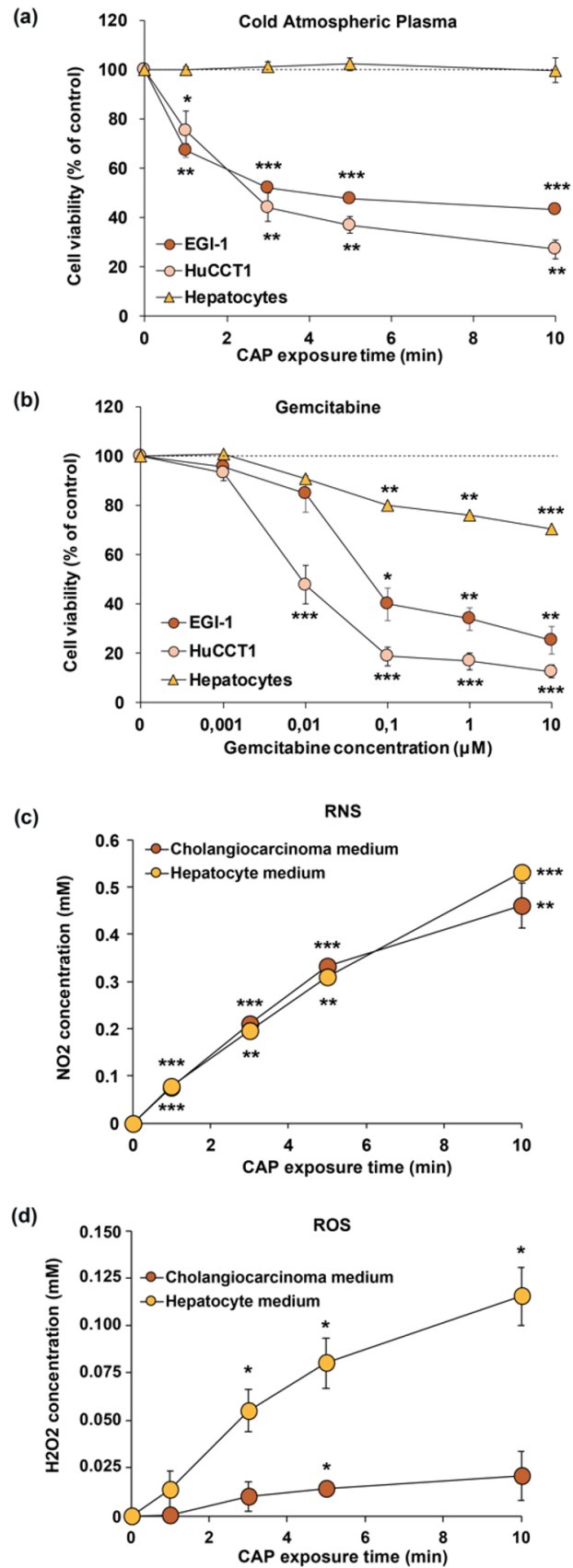

Next, we performed *in vitro* studies on CCA cell lines to further dissect the effects induced by CAP on tumor cells. First, we evaluated the effects of CAP treatment on the viability of two human CCA cell lines, EGI-1, the same cell line used for the induction of subcutaneous xenografts, and HuCCT1. Besides, to verify if CAP treatment is biologically selective, non-malignant primary human hepatocytes, the main cell type in the liver, where isolated from patients. They were also exposed to the same CAP treatment to verify if CAP may drive to side effects. To standardize the application of CAP across the different *in vitro* experiments, first we treated by plasma a standard volume of fresh culture media (3 mL) in a standardized plastic support (6-well plates) for 3 min. Second, we incubated the resulting plasma-activated culture media (commonly called PAM) with either CCA cell lines or human hepatocytes in culture (Supplementary Figure 1). Such indirect CAP treatment induced a decrease in the viability of CCA cells and this effect became stronger for CAP exposure times increasing from 1 to 10 minutes (Figure 4a). In contrast, no effect was observed on the viability of human hepatocytes isolated from 3 different patients (Figure 4a), hence demonstrating a selective effect of CAP on tumor cells over non-malignant liver cells. Of note, similar experiments performed after exposure to gemcitabine showed a dose-dependent decrease in cell viability that was more pronounced in CCA cells but reached approximately a 30% reduction in hepatocytes (Figure 4b), demonstrating a better selectivity of CAP over gemcitabine. Since CCA cell lines and primary hepatocytes need different culture media due to specific requirements of each cell type, we evaluated the production of RONS in media. More specifically, we determined the concentration of $NO_2$ and $H_2O_2$ in CAP-exposed culture media at different time points, the same used in cell viability studies. While the production of $NO_2$ remains overall the same over treatment time in both types of media (Figure 4c), production of $H_2O_2$, was approximately 6 times higher in hepatocyte media than in CCA media (Figure 4d). To get more insight on this issue we determined the generation of ROS in cell lysates from CCA cells and hepatocytes exposed to PAM. Interestingly, production of $H_2O_2$ was only increased in CCA cells exposed to PAM, while it remained unchanged in hepatocytes (Figure 4e). This observation led us to think about potential defense mechanisms protecting hepatocytes from ROS production, more specifically, ROS-scavenging enzymes. Indeed, further analysis revealed that mRNA expression of several enzymes was strongly increased in hepatocytes compared to both CCA cell lines (Figure 4f). Altogether, these results validate the selective effect of CAP-activated medium in CCA cells over hepatocytes.

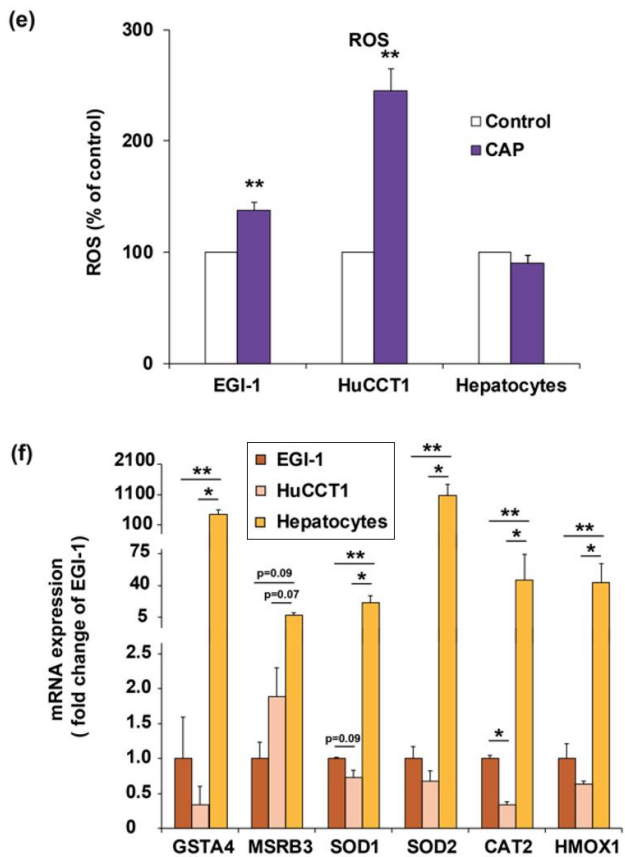

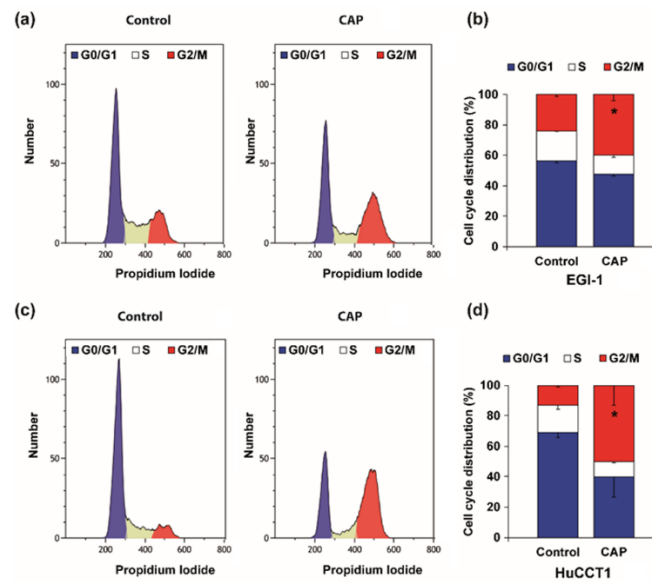

*Figure 4. (a-b) Effect of CAP (a) and gemcitabine (b) on the viability of EGI-1 and HuCCT1 CCA cells and human primary hepatocytes. Cell viability was measured after incubation for 72 hours with culture medium previously treated for 1, 3, 5 and 10 min with CAP (9kV, 30kHz, 14%, gap of 7 mm). (c-d) $NO_2$ (c) and $H_2O_2$ (d) determination in culture media from CCA cells and primary hepatocytes. (e) $H_2O_2$ determination in cell lysates from CCA cells and primary hepatocytes exposed to PAM for 3 min. (f) Expression of GSTA4, MSRB3, SOD1, SOD2, CAT2 and HMOX1 at mRNA level in CCA cell and hepatocytes. Values are expressed as means ± SEM from at least 3 independent cultures. \*, p < 0.05; \*\*, p < 0.01; \*\*\*, p < 0.001; compared with untreated cells (0 min).*

## II.4. Cold atmospheric plasma induces cell cycle arrest and apoptosis in cholangio-carcinoma cells

As previously underlined, CAP-derived RONS drive cell cycle arrest and cell death by damaging DNA [6,7]. For the following experiments we used the IC50 from the viability assays (Figure 4a), corresponding to 3-min treatment with CAP. Therefore, we evaluated the possibility of cell cycle arrest in our experimental conditions. Indeed, flow cytometry analysis of cell cycle distribution showed changes in the different phases (Figure 5). Both EGI-1 and HuCCT1 cells experienced a decrease in the percentage of cells in G0/G1 phases and S, and an increase of the percentage of cells in G2/M phases.

*Figure 5. (a-d) Representative flow cytometry cell cycle measurement (a, c) and graphical representation of the cell cycle distribution (b, d) of EGI-1 (a-b) and HuCCT1 (c-d) CCA cells after 24h of exposure to culture medium pretreated with CAP for 3 min (9kV, 30kHz, 14%, gap of 7 mm). Cell populations in G0/G1, S, and G2/M phases are given as percentage of total cells. Values are expressed as means ± SEM from at least 3 independent cultures. \*, p < 0.05; compared with control cells.*

The accumulation of cells in G2/M indicate that cells arrested the cell cycle at the G2/M DNA damage checkpoint, that serves to prevent cells with genomic DNA damage from entering the M phase. Therefore, our next step was to determine if, as observed *in vivo*, CAP treatment could drive DNA damage in CCA cells *in vitro*. One of the most important proteins required for checkpoint-mediated cell cycle arrest and DNA repair following double-stranded DNA breaks is the histone H2AX. DNA damage caused by oxidative stress results in a rapid phosphorylation of H2AX (named ᵧH2AX), leading to the recruitment of several proteins in response to DNA damage. Immunofluorescence analysis showed a strong staining of phospho-histone H2AX in both EGI-1 and HuCCT1 cells at different times (*i.e.* 24h, 48h and 72h) after exposure to CAP-activated culture medium compared to untreated cells (Figure 6a and d), being 72h in EGI-1 and 48h in HuCCT1 cells, the highest signal, as ascertained by western blot (Figure 6b-c and e-f). Western blot analyses showed a clear correlation between the increase of histone H2AX phosphorylation and PARP cleavage (Figure 6b-c and d-e), a marker of cell apoptosis.

To better decipher the mechanism of cell cycle arrest in CCA cells treated with CAP, we evaluated the activation of the two parallel signaling pathways that ultimately break the cell cycle once the DNA damage is sensed. These signaling cascades that block the progression to mitosis are led by CHK kinases and p53, respectively. Western blot analysis from Figure 7b-c and e-f showed a strong phosphorylation of both CHK1 and p53 from 24h to 72h in both cell lines. These results suggest that the cell cycle is arrested soon after CAP-activated culture medium exposure, when DNA damage is first detected, but apoptosis is not induced until the accumulation of DNA damage is strong enough, that is 72h after exposure to CAP in EGI-1 and 48h in HuCCT1 cells. Interestingly, CAP exposure of hepatocytes showed a reduced expression of CHK1 and p53 compared to CCA cells (Supplementary Figure 2 a), probably due to the low proliferative capacity of these cells in primary culture. Additionally, no changes in H2AX phosphorylation or PARP cleavage were observed, indicating the absence of DNA damage and corroborating the selective capacity of CAP (Supplementary Figure 2 a).

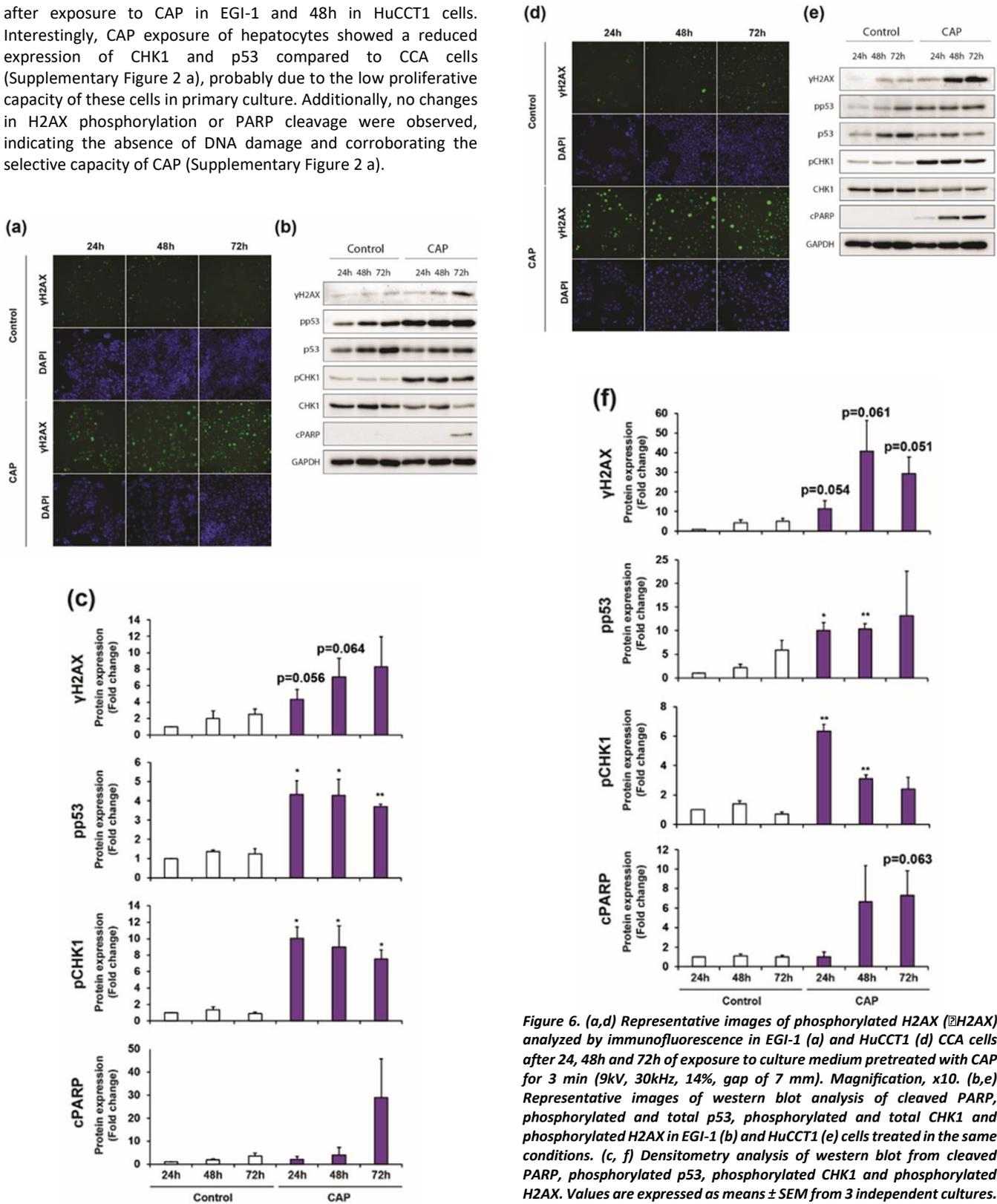

*Figure 6. (a,d) Representative images of phosphorylated H2AX (γH2AX) analyzed by immunofluorescence in EGI-1 (a) and HuCCT1 (d) CCA cells after 24, 48h and 72h of exposure to culture medium pretreated with CAP for 3 min (9kV, 30kHz, 14%, gap of 7 mm). Magnification, x10. (b,e) Representative images of western blot analysis of cleaved PARP, phosphorylated and total p53, phosphorylated and total CHK1 and phosphorylated H2AX in EGI-1 (b) and HuCCT1 (e) cells treated in the same conditions. (c, f) Densitometry analysis of western blot from cleaved PARP, phosphorylated p53, phosphorylated CHK1 and phosphorylated H2AX. Values are expressed as means ± SEM from 3 independent cultures. \*, $p < 0.05$; \*\*, $p < 0.01$; compared with control cells.*

When these experiments were reproduced after exposure to gemcitabine, we observed similar results in terms of increase of H2AX, CHK1 and p53 phosphorylation, accompanied by PARP cleavage in both CCA cell lines (Supplementary Figure 2b). Interestingly, gemcitabine induces DNA damage in hepatocytes in a dose dependent manner (Supplementary Figure 2c), concordant with the decrease in viability observed in Figure 4b, and this DNA damage started as early as 24h after exposure and was maintained until 72h, as ascertained by H2AX phosphorylation (Supplementary Figure 2d). However, no change was observed in the phosphorylation of CHK1 and p53 or PARP cleavage, indicating that the reduction in hepatocyte viability induced by gemcitabine may not be related to cell cycle arrest and apoptosis, but other types of dead such as necrosis or senescence.

Finally, we verified that the decrease in cell viability of EGI-1 and HuCCT1 after CAP treatment was due to apoptosis. Indeed, exposure of cells to PAM reduce the number of viable cells and increase the populations in the quadrants corresponding to late-apoptotic and necrotic cells in both cell types (Figure 7a-d), as ascertained by Annexin V-7AAD quantification by flow cytometry. Of note, this increase in apoptotic cells was observed in HuCCT1 at 48h, but it was not in EGI-1 at this time (data not shown), becoming only evident at 72 h in the later. These results may corroborate that apoptosis is not induced until the accumulation of DNA damage is strong enough, that is 72h after exposure to PAM in EGI-1 and 48h in HuCCT1 cells.

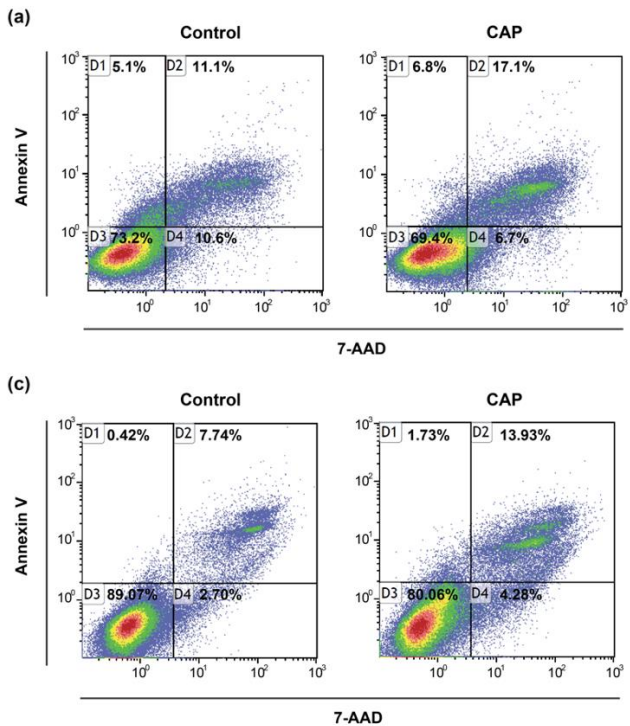

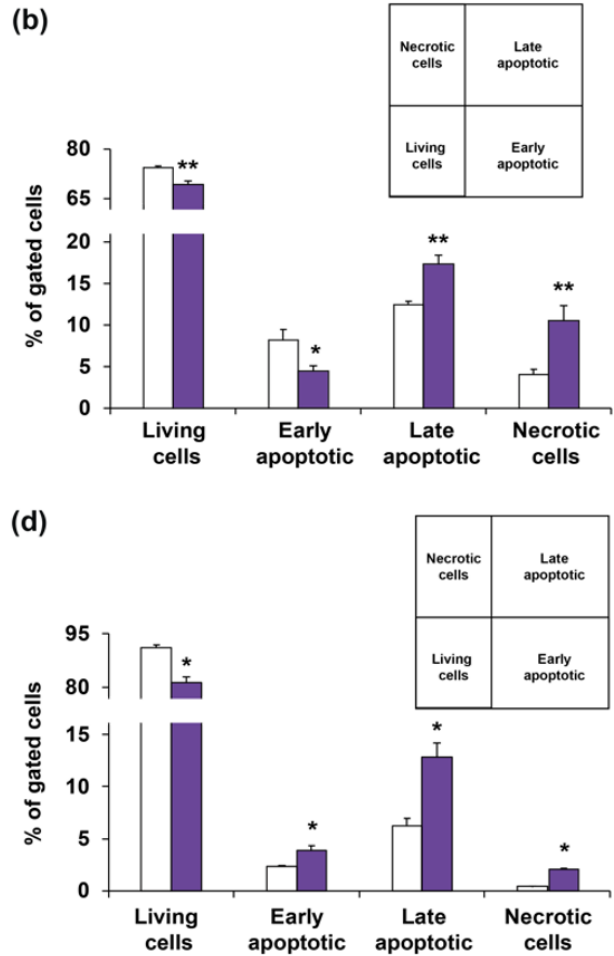

*Figure 7. (a-d) Representative images (a, c) and quantification (b, d) of apoptosis by flow cytometry analysis of Annexin V/7AAD in EGI-1 (a-b) and HuCCT1 (c-d) CCA cells after 48 h of exposure to PAM for 3 min (9kV, 30kHz, 14%, gap of 7 mm). Values are expressed as means ± SEM from at least 3 independent cultures. *, p < 0.05; **, p < 0.01; compared with control condition.*

## II.5. Cold atmospheric plasma affects the phenotype of tumor-associated macrophage

Besides effects of CAP on tumor cells, we sought to determine if CAP exposure may have any effect on the stroma of the EGI-1 subcutaneous xenograft model. This model has the advantage of providing the opportunity of evaluating the expression of human genes, corresponding to the injected tumor CCA cells, and murine genes, corresponding to the cells forming the stroma that are recruited cancer cells during tumor formation. Therefore, we examined the mRNA expression of different specific markers corresponding to cancer-associated fibroblasts (CAF) (Acta2, coding alpha-SMA), endothelial cells (EC) (Pecam1, coding for CD31) and tumor-associated macrophages (TAM) (Adgre1, coding for F4/80). There were no significant changes in the mRNA of Acta2 or Pecam1 among the different groups (Figure 8a). However, the expression of Adgre1 increased in the tumors from the animals that received CAP or gemcitabine treatment compared to the controls, suggesting a potential enhanced recruitment and/or proliferation of TAMs in the treated tumors (Figure 8a). Presence of TAM in tumor from the different groups was evidenced by immunohistochemical analyses of F4/80, as shown in representative images from each group (Figure 8b), although it was impossible to properly determine differences in macrophage infiltration by F4/80 IHC quantification. However, based on previous publications indicating a phenotypic change of macrophages in absence of changes in the total number of these cells after exposure to experimental therapies [11], we decided to perform a preliminary analysis to elucidate this point. Analysis of Ccl2 (coding for Monocyte chemotactic protein-1, MCP-1) and Ccr2, a chemokine and its receptor, respectively, which are major regulators of monocyte chemotaxis and macrophage trafficking, showed an increased expression in groups treated with CAP and gemcitabine compared to the controls (Figure 8c), which may suggest changes in chemotactic response of resident TAM. In addition, CAP was able to increase the expression of several cytokines that are associated with the antitumor phenotype of macrophages and that are involved in the induction of apoptosis, i.e. Tnfa (coding for Tnfα), Tnfsf1 (coding for TNF-related apoptosis-inducing ligand (Trail)) and Il1b (coding for Il1β) (Figure 8d). These results are in accord with previous publications that link CAP treatment with modulation of immune cells and together with the increasing interest of immunotherapies as cancer treatment validate the need for further investigation on this topic in CCA.

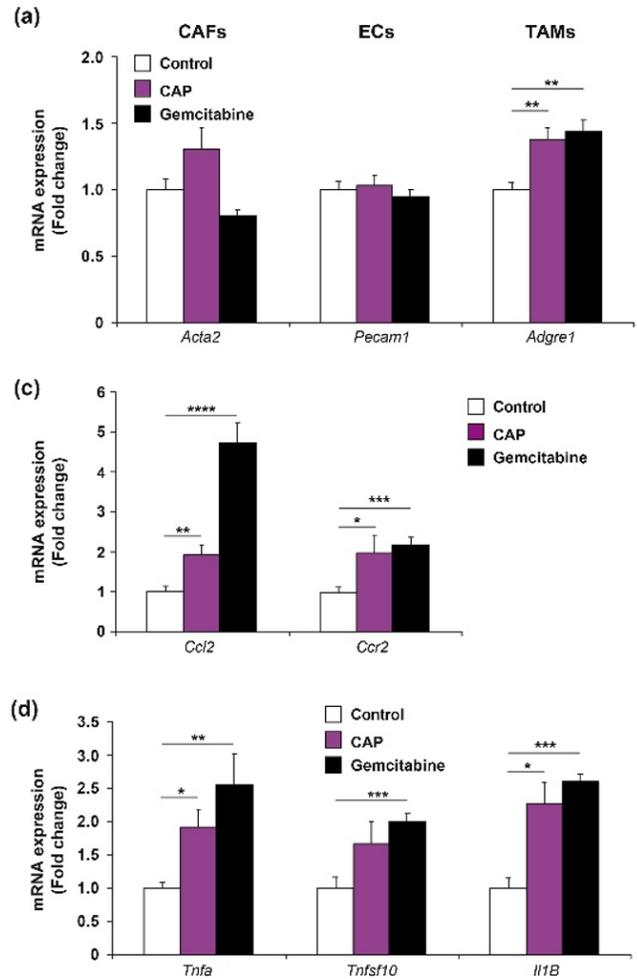

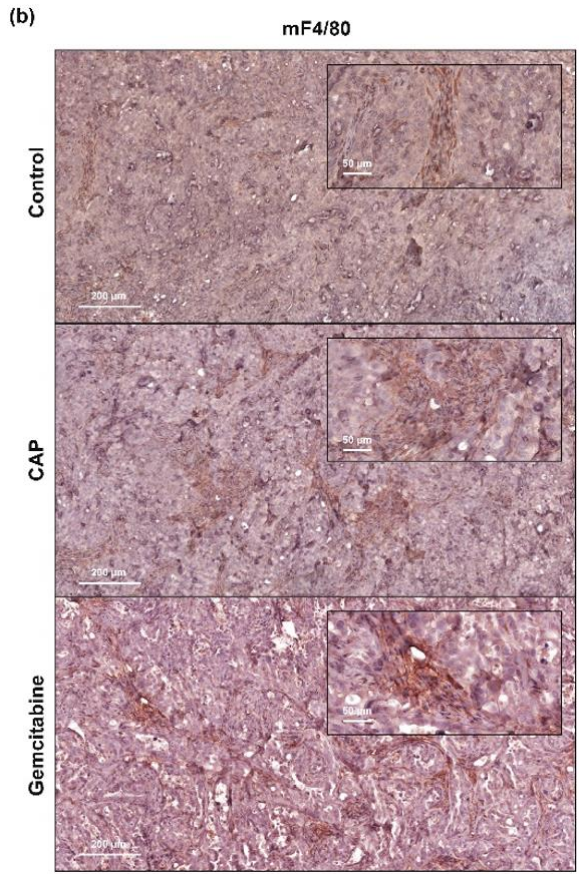

*Figure 8. (a) Changes in mRNA expression of cell type markers (Acta2/α-SMA, a marker of cancer-associated fibroblasts CAFs, Pecam1/CD31, a marker of endothelial cells (EC) and Adgre1/F4/80, a marker of tumor-associated macrophages (TAM) in control (white bars), CAP (purple bars) and gemcitabine (black bars) treated xenograft tumors. (c) Representative IHC staining of F4/80 in the same tumors. Magnification x250 (inserts x1000). Scale: 200 μm. (b) Changes in mRNA expression of Ccl2/Mcp1 and Ccr2 (c) in control (white bars), CAP (purple bars) and gemcitabine (black bars) treated xenograft tumors. (d) Changes in mRNA expression of pro-apoptotic cytokines (Tnfa/Tnfα, Tnfsf1/Trail and Il1b/Il1β) in control (white bars), CAP (purple bars) and gemcitabine (black bars) treated xenograft tumors. Values are expressed as means ± SEM. \*, $p < 0.05$; \*\*, $p < 0.01$; \*\*\*, $p < 0.001$; compared with control tumors.*

## III. Discussion

In the present work, we analyzed the effects of CAP *in vivo* in a mouse xenograft model of CCA and *in vitro* on human CCA cell lines as well as on non-malignant human hepatocytes. We found that local application of CAP on the tumor halts its growth without inducing systemic side effects. Analysis of tumors showed areas of calcification suggesting cell dead, that was confirmed by immunostaining of cleaved-caspase-3, a protein of the apoptotic pathway, along with DNA lesions due to plasma-originated reactive species. *In vitro*, CAP-activated medium contains reactive species (e.g. nitrites) that induced oxidative stress and reduced cell survival by arresting the cell cycle and inducing apoptosis in CCA cells but not in hepatocytes. Finally, preliminary analysis suggested changes in the surrounding stroma of CCA tumors after exposure to CAP.

Since the early 2000s, CAP have generated a lot of interest in cancer medicine as a promising treatment for cancer without inducing systemic toxic side effect. Anti-tumor properties of CAP are now well established and tumor volume reductions have been demonstrated in murine tumor models of several cancer types including pancreatic [12,13], ovary [14], breast [15] and colon [16], melanoma [17] and glioblastoma [6,18]. Since CCA is a very aggressive tumor with a limited therapeutic arsenal, we investigated if CAP may drive to anti-cancer effects *in vivo*. Based on our previous work that aimed to set up a safe device with anti-tumor properties in CCA [9], we conducted further studies to decipher in a deeper way the cellular mechanisms behind CAP effect. Up to date the only two studies dealing with the effects of CAP on liver cancer were performed in hepatocellular carcinoma cell lines [5,8]. Thus, this study is the first conducted on CCA using *in vivo* and *in vitro* preclinical models.

Owing to the emerging and highly multidisciplinary aspects of "cold plasma oncology", only 5% of the studies published so far include *in vivo* experiments [19]. Most CAP studies in cancer have been achieved using tumor cell lines originating from either solid or blood tumors and rarely on mouse tumor models. We conducted *in vivo* studies to analyze the effects of CAP on death and oxidative stress, and we compared this treatment to a conventional treatment with gemcitabine. In our study, CAP demonstrated anti-tumor properties although a traditional chemotherapeutic agent such as gemcitabine showed higher efficiency. Interestingly, CAP was applied locally on a very small tumor surface for a very short period of time (1 minute) demonstrating no side effects, while gemcitabine, that was applied intraperitoneally, was accompanied by an increased plasmatic concentration of markers indicating liver damage. Even if few studies have been performed *in vivo*, some of them confirmed that CAP has no systemic effects. Liedtle et al. have addressed this point through a complete study showing that CAP by using plasma-activated medium does not affect blood parameters, leucocyte distribution or cytokine signature [20]. However, classical blood parameters to evaluate liver and cell toxicity, such as transaminases and LDH, were not measured, in contrast to our study. In spite of our in vitro observation on primary hepatocytes and the absence of liver damage *in vivo*, studies using orthotopic CCA model are required to evaluate the direct effect of CAP on liver parenchyma. Nevertheless, further investigation to improve the surface exposure and the time of treatment with CAP is crucial to obtain the maximum benefit from this new therapeutic tool.

At cellular level, histology examination of the tumor showed signs of calcification, a reaction occurring in response to cell injury, indicating the presence of apoptotic tissue. Activation of signaling pathways involved in cell death was confirmed by the immunohistochemical analysis of cleaved caspase-3 suggesting an induction of caspase-3-dependent apoptosis in tumor cells. Induction of cell apoptosis is the primary mechanism of CAP action following the reactive species generated by CAP [19]. However, other cell death pathways have been recently evidenced such as ferroptosis in tumor cells subjected to CAP treatment [21]. In CCA cell lines, we tested cell media that were first treated by CAP, i.e. PAM. Then PAM was immediately transferred to the cell culture. Indirect or direct treatment by CAP displays similar efficacy on tumor cell culture, and PAM is also able to reduce tumor burden without inducing side effects when injected intraperitoneally in a murine model of pancreatic cancer [20]. Intraperitoneal injection of PAM lead to reduced metastatic potential of ovarian and gastric cancer cells [22,23]. When we evaluated PAM on CCA cells, although PAM decreased cell survival in both CCA cell lines with similar efficacy, induction of apoptosis was lower in EGI-1 than in HuCCT1 cells. Doses of CAP used to treat the medium matters and as suggested in previous studies, low doses of CAP can inhibit cell proliferation without inducing apoptosis, but instead induce senescence [24,25] or autophagy [26,27]. In addition, CAP can affect other cell biology features, for example by inducing endoplasmic reticulum stress, depolarization of mitochondrial membrane potential, DNA damage or by decreasing migratory and invasive properties [22,28,29], although these aspects deserve further characterization in CCA

At molecular level, we detected DNA double strand breaks in both CCA cell lines, along with DNA damage responses with an upregulation of the phosphorylation status of p53 and of CHK1, both regulating cell cycle checkpoints. We previously observed similar DNA damage in CCA cells subjected to oxidative stress with hydrogen peroxide [30], suggesting that upon CAP treatment, CCA cells may undergo oxidative stress. Overload of RONS in CCA cells leads to DNA damage, attested by the phosphorylation of histone H2AX, triggering pathways that will ultimately kill the cancer cells [31]. Altogether, these results fit perfectly with previous finding in other tumors, such as oral cancer, were p53 signaling pathway was identified as one of the most deregulated pathways after exposure to PAM by using RNA-sequencing approaches [32].

Targeting specifically tumor cells without damaging healthy cells is a major challenge of anti-cancer treatment. CAP has the advantage to selectively induce cell cycle arrest and death of tumor cells but not of healthy ones. Whatever the direct/indirect approach, the concept of plasma selectivity is a key issue in treatment. Pioneering studies from Babington *et al.* have shown that the plasma treatment of mice bearing subcutaneous glioblastoma led to a 56% decrease of tumor volume while maintaining the viability of healthy cells surrounding the tumor at 85% [33]. While CAP had a significant effect on CCA cancer cells by decreasing cell viability, it had no deleterious effect on non-malignant liver cells, i.e. primary human hepatocytes, suggesting a selectivity of CAP treatment. By killing primarily cancer cells, plasma treatment preserves healthy tissue and thereby tissue function. Keidar et al were amongst the first to demonstrate a selectivity of CAP on lung cancer cell lines *versus* normal human bronchial epithelial cells [34]. This selectivity was also emphasized in melanoma cells compared to normal keratinocytes [35], and other cancer types (ovarian, glioblastoma), as a general property of CAP [36]. However, all these studies deal with cell lines but none with primary cells. In our studies, hepatocytes were isolated from human liver and cultured according to a well-defined protocol [37]. We found that CAP has no impact on hepatocyte survival nor the induction of DNA damage or apoptotic regulatory signaling pathways, in contrast to CCA cell lines. We chose hepatocytes as non-tumor cells because they are the most abundant cell type of the liver. Although the media composition, an essential parameter [36], was not the same between the two cell types, CAP generated the same profile of RNS in both media and higher ROS in hepatocyte media. Furthermore, hydrogen peroxide increased in CCA cell lines after exposure to PAM, as previously described for atmospheric pressure plasma jets [38], while it remained unchanged in hepatocytes. The cellular mechanisms by which CAP operates this selectivity are still poorly understood and indirect evidence exist to explain this crucial issue. Among the potential mechanisms given so far aquaporins and anti-oxidant cellular defense systems seem to be the most plausible explanations [4]. Indeed, as happened in our study, elevated expression of ROS-scavenging enzymes such as superoxide dismutase, catalase and glutathione reductase has been observed in healthy cells compared to tumor cells, which may contribute to cellular defense against CAP-originated reactive species [4].

Finally, one major point that should be considered when a tumor is treated by CAP is its potential effect on the tumor microenvironment cells. Tumor is a mix of several cell types including tumor cells but also CAF, EC and TAM. According to histological examination of CCA tumors treated with CAP, fibrotic stroma is not affected by CAP treatment, a result that is confirmed by unchanged mRNA expression level of a-SMA, a marker of CAF, between treated and untreated conditions. As previously shown, fibroblasts are less affected by CAP compared to cancer cells [20,39]. No obvious change in vascularization is observed even if plasma has been shown to suppress neovascularization but not pre-existing vessels, an effect that is partly independent of ROS [40]. Further studies must be conducted in the case of CCA to confirm or not a potential action of CAP on vascular system. Interestingly, one of the most promising views is that CAP treatment is able to activate the immune response to attack the tumor [16,41,42]. Indeed, our analysis on tumor xenografts showed changes in the expression of markers related to TAM phenotype, suggesting a potential shift towards an anti-tumor phenotype of TAM, although this issue deserves further consideration and new research will be undertaken. *In vitro* studies performed by other groups are in agreement with our findings *in vivo*, suggesting that increasing the function of pro-inflammatory macrophages may help to control tumorigenesis caused by compromised immune response [41]. Taking into account that one of the most therapeutic strategies under study nowadays is the activation of the patient immune system to fight tumors, it is imperative to keep deepening on the molecular mechanisms implicated in the effects of CAP on the immune system, especially in immunocompetent murine cancer models in which not only macrophages but also lymphocytes could be potentially involved in this response.

## IV. Materials and Methods

### IV.1. Cell culture and treatment

HuCCT1 cells, derived from intrahepatic biliary tract, were kindly provided by Dr. G. Gores (Mayo Clinic, MN). EGI-1 cells, derived from extrahepatic biliary tract, were obtained from the German Collection of Microorganisms and Cell Cultures (DSMZ, Germany). Cells were cultured in DMEM supplemented with 1 g/L glucose, 10 mmol/L HEPES, 10% fetal bovine serum (FBS), antibiotics (100 UI/mL penicillin and 100 mg/mL streptomycin), and antimycotic (0.25 mg/mL amphotericin B). Cell lines were routinely screened for the presence of mycoplasma and authenticated for polymorphic markers to prevent cross-contamination.

### IV.2. Isolation and culture of human hepatocytes

Normal liver tissue was obtained from adult patients undergoing partial hepatectomy for the treatment of colorectal cancer metastases. Primary human hepatocyte isolation was performed on the Human HepCell platform (IHU-ICAN, Paris, France, http://www.ican-institute.org/category/plateformes) as previously described [43]. Ethical approval for the isolation of human hepatocytes was granted by the Persons Protection Committee (CPP Ile de France III) and by the French Ministry of Health (N°: COL 2929 and COL 2930). Hepatocytes were isolated using an established two-step-perfusion protocol with collagenase. First, the tissue was rinsed with pre-warmed (37 °C) calcium-free buffer supplemented with 5 mmol/L ethylene glycol tetraacetic acid (Sigma, Saint-Quentin Fallavier, France). Then, the liver sample was perfused with recirculating perfusion solution containing 5 mg/mL of collagenase (Sigma) at 37 °C. Afterwards, the tissue was transferred into a petri dish containing a Hepatocyte Wash Medium (Life technologies, Villebon sur Yvette, France). Tissue was disrupted mechanically by shaking and using tweezers to disrupt cells from the remaining scaffold structures. Cellular suspension was filtered through a gauze-lined funnel. Cells were centrifuged at low speed centrifugation (50 g). The supernatant was removed, and pelleted hepatocytes were re-suspended in Hepatocyte Wash Medium. Viability cell was determined by trypan blue exclusion test. Freshly isolated normal hepatocytes were suspended in Williams' medium E (Life Technologies) containing 10% fetal calf serum (FCS) (Eurobio, Courtaboeuf, France) penicillin-streptomycin (penicillin: 200 U/mL; streptomycin: 200 µg/mL) and insulin (0,1 U/mL). Then, the cells were seeded in 6 and 96 well plates pre-coated with type I collagen at a density of 1.8 x$10^6$ and 0.5 x$10^5$ viable cells/well, respectively, and incubated at 37°C in a 5% $CO_2$ overnight. Then, the medium was replaced with fresh complete hepatocyte medium supplemented with 1 µmol/L hydrocortisone hemisuccinate (SERB, Paris, France) and cells were left in this medium until treatment with plasma activated medium (PAM).

### IV.3. Xenograft tumor model

Animal experiments were performed in accordance with the French Animal Research Committee guidelines and all procedures approved by a local ethic committee (No 10609). 2 × $10^6$ of EGI-1 cells were suspended in 60 µL of PBS and 60 µL of Matrigel® growth factor reduced (Corning) and implanted subcutaneously into the flank of 5-week-old female ATHYM-Foxn1 nu/nu mice (Janvier Labs, France). Mice were housed under standard conditions in individually ventilated cages enriched with a nesting material and kept at 22 °C on a 12 h light/12 h dark cycle with ad libitum access to food and tap water. Tumor growth was monitored by measuring every 2–3 days the tumor volume (V xenograft) with a caliper as follows: V xenograft = x·$y^2$/2 where x and y are the longest and shortest lateral diameters respectively. Once tumor volume reached approximately 200 $mm^3$, CAP and gemcitabine treatments were initiated. Gemcitabine was administered every Monday and Thursday during 3 weeks by intraperitoneal injection at a concentration of 120 mg/kg dissolved in saline solution (vehicle). Cold atmospheric plasma was administered as explained in section 4.4 the same days as gemcitabine.

### IV.4. Cold atmospheric plasma treatment

The *in vivo* and the *in vitro* experiments were conducted using the same atmospheric pressure plasma jet device, called PTJ, as sketched in Figure 1a. It is composed of a 10 cm long dielectric quartz tube presenting a 4 mm inner diameter and a 2 mm wall thickness. Its electrode configuration is made of two outer ring electrodes with inner and outer diameters of 8mm and 12.8 mm respectively while the inter-ring distance is 50 mm. For all experiments, the lower ring electrode was connected to the ground while the upper ring electrode was biased to the high voltage. The PTJ was supplied with helium gas (flow rate of 1 slm) and powered with a nanopulse high voltage generator device (model Nanogen 1) from RLC Electronic company. For both *in vivo* and *in vitro* experiments, electrical parameters were fixed as follows: 9 kV of amplitude, 14% of duty cycle and 30 kHz of repetition frequency. The reasons explaining how these values were chosen as well as the physico-chemical characterizations of the PTJ device have already been published in [9]. For the *in vivo* studies, the cold atmospheric plasma was applied to the animals as previously described [9], 9kV, 30kHz, 14%, maintaining a gap of 10 mm between the tube and the skin. For the *in vitro* studies, cells were treated with PAM. In order to maintain reproducibility among different plastic supports, 3 mL of the corresponding culture media in a 6-well plate were treated with the same conditions (9kV, 30kHz, 14%) during 1, 3, 5 or 10 min. A gap of 7 mm between the tube and the surface of medium was constantly maintained. After treatments, PAM was transferred to 96-, 24- or 6-well plates according to the different analysis performed (Supplementary Figure 1).

### IV.5. Biochemistry

The concentrations of alanine aminotransferase (ALAT), aspartate aminotransferase (ASAT), and lactate dehydrogenase (LDH) in plasma of mice, were measured on an Olympus AU400 Analyzer.

## IV.6. Histology and (immuno)histochemistry

Formalin-fixed paraffin-embedded tissue samples from mice xenografts were cut in 4 μm sections, deparaffined and stained with hematoxylin and eosin to observe tissue histology.
For immunohistochemistry, antigens were unmasked as indicated in Table 1. For cleaved-caspase-3 and 8-oxoguanine, sections were sequentially incubated with $H_2O_2$ for 5 minutes (only for caspase3), with Protein Block (Novolink Polymer Detection System; Novocastra Laboratories Ltd.) for 5 minutes, and with primary antibodies for 30 minutes (overnight for 8-oxoguanine). Novolink Post Primary was applied for 15 minutes. Sections were finally washed and incubated with Novolink Polymer for 15 minutes. An automated staining system (Autostainer Plus, Dakocytomation) was used to perform immunostaining. The color was developed using amino-ethyl-carbazole (AEC peroxidase substrate kit; Vector Laboratories). Sections were counterstained with hematoxylin and mounted with glycergel (Dako). For F4/80 sections were incubated with PBS 0.5% triton X-100 30 minutes to increase the permeabilization of the tissue. Then, they were blocked with horse serum 2.5% (Vector) during 1h. After tissue blocking, samples were immunostained with primary antibody overnight at 4°C. Then, endogenous peroxidase blocking was performed with hydrogen peroxide solution (Leica) during 1h. Samples were developed with the ImPRESS Excel staining kit (Vector) following manufacturer instruction. Briefly, tissue samples were incubated with anti-rabbit Ig secondary antibody for 1h30, washed with PBS, and incubated with an anti-goat amplifier antibody for 1h. Finally, the samples were developed with peroxidase substrate for 3 minutes and counterstain with Mayer's hematoxylin (Dako) 5 minutes.

**Table 1. Primary antibodies used for immunodetection.**

| Name | Species | Manu-facturer | Reference | Dilution | Antigen unmasking |
|---|---|---|---|---|---|
| 8-oxoguanine | M | Abcam | ab206461 | 1/100 (IHC) | EDTA pH8 |
| cCaspase3 | R | CST | CST9664 | 1/100 (IHC) | Citrate pH6 |
| cPARP | R | CST | CST5625 | 1/1000 (WB) | |
| CHK1 | M | CST | CST2360 | 1/1000 (WB) | |
| pCHK1 | R | CST | CST2348 | 1/1000 (WB) | |
| F4/80 | R | Spring Bioscience | M4154 | 1/100 (IHC) | Citrate pH6 |
| GAPDH | M | Santa Cruz | sc-32233 | 1/5000 (WB) | |
| p53 | M | Santa Cruz | sc-126 | 1/500 (WB) | |
| pp53 | R | CST | CST9284 | 1/1000 (WB) | |
| γH2A.X | R | CST | CST9718 | 1/1000 (WB), 1/200 (IF) | |

M, mouse; R, rabbit; WB, western blot; IF, immunofluorescence; IHC, immunohistochemistry.

## IV.7. Cell viability

5000 EGI-1 cells/well, 4000 HuCCT1 cells/well and 50,000 hepatocytes/well were plated in 96-well plates. 24 h later, the medium was replaced by fresh culture medium, PAM or gemcitabine. Cells were then incubated for 72 hours before determining the viability by the crystal violet method. Absorbance was quantified with a spectrophotometer (Tecan) at 595 nm.

## IV.8. RONS determination in culture media

To verify whether reactive species are produced in PAM, nitrites and $H_2O_2$ concentrations were measured using Griess reagent (Sigma Aldrich) and Titanium Sulfate TiSO4 (Sigma-Aldrich) respectively. In presence of nitrite species, the Griess reagent shows an absorption peak at 518nm (pink coloration) while in presence of peroxide, the TiSO4 shows an absorption peak at 405nm (yellow coloration) both measured with the Biotek Cytation 3 device. A two-steps protocol was followed: first, the media were placed in 6 well-plates and exposed to plasma as previously explained in section 4.4. Second, plasma was switched off. For the nitrite determination, 25 mL of each culture media sample was mixed with 175 mL of distilled water and 50 mL of Griess reagent. For the peroxide determination, 250 mL of each culture media sample was mixed with 100 mL of $TiSO_4$.

## IV.9. ROS determination in cell lysates

ROS production was assessed using the 2',7'-dichlorofluorescein diacetate (H2DCFDA; Abcam cat number ab113851) according to the instructions. Briefly, CCA cells and hepatocytes were plated at 2.5 x$10^5$ cells/well and 0.5 x$10^5$ cells/well, respectively, in black-walled, clear-bottom 96-well microplates, and incubated for 24 h at 37 °C. The cells were incubated with CM-H2DCFDA (25 μM) in PBS for 30 min and then with PAM for 30 min. Cells were washed with PBS, and fluorescence was measured at 485/535nm (Tecan). Normalization was done by crystal violet method.

## IV.10. Apoptosis assay

2x$10^5$ EGI-1 cells/well and 1.5x$10^5$ HuCCT1 cells/well were plated in 6-well plates. 24 h later, the medium was replaced by fresh culture medium or PAM. 48h or 72h later both cells from the supernatant and the plates were collected and stained using the PE Annexin V Apoptosis Detection Kit with 7-AAD (BioLegend), according to the manufacturer's instructions. Flow-cytometric analysis was performed using a Gallios flow cytometer (Beckman-Coulter) to calculate the apoptosis rate. Results were analyzed using Kaluza analysis software (Beckman-Coulter).

## IV.11. Immunofluorescence

Immunofluorescence assays were performed as previously described [44]. Primary antibodies are provided in Table 1. Cells were observed with an Olympus Bx 61 microscope (Olympus).

## IV.12. Western blot analysis

For obtaining whole-cell lysates for WB, cell cultures were lysed in RIPA buffer supplemented with 1 mmol/L orthovanadate and a cocktail of protease inhibitors. Proteins were quantified using a BCA kit (Pierce). WB analyses were performed as previously described [44]. Primary antibodies are provided in Table 1.

## IV.13. Cell cycle analysis

$0.6 \times 10^5$ EGI-1 cells/well and $0.5 \times 10^5$ HuCCT1 cells/well were seeded in 6-well plates and incubated for 24 h. The cells were then treated with PAM for 24h. Cells are detached with trypsin, washed with cold PBS, pooled, and centrifuged before being fixed in 70% ice-cold ethanol during 30 min at -20°C, and stored at −20 °C if required. Cells are incubated with 100 µg/mL of RNase A and 40 µg/mL of propidium iodide in PBS buffer. The stained cells were analyzed with a CytoFLEX (Beckman-Coulter), and their distribution in different phases of the cell cycle was calculated using Kaluza analysis software.

## IV.14. RNA and reverse transcription-PCR

Total RNA extraction and RT-qPCR was performed as previously described [44]. Primer sequences are provided in Table 2. Gene expression was normalized to Hprt1 mRNA content for mouse genes and was expressed relatively to the control condition of each experiment. The relative expression of each target gene was determined from replicate samples using the formula $2^{-\Delta\Delta Ct}$.

**Table 2. Mouse primer used for quantitative real-time PCR.**

| Gene | Protein | Forward (5'→3') | Reverse (5'→3') |
|---|---|---|---|
| Acta2 | α-Sma | CTGTCAGGAACCCTGAGACGCT | TACTCCCTGATGTCTGGGAC |
| Pecam1 | CD31 | AGCCTCCAGGCTGAGGAAAA | GATGTCCACAAGGCACTCCA |
| Ccr2 | Ccr2 | GGCCACCACACCGTATGACTA | AGAGATGGCCAAGTTGAGCAGATAG |
| Adgre1 | F4-80 | CTTTGGCTATGGGCTTCCAGTC | GCAAGGAGGACAGAGTTTATCGTG |
| Il1b | Il1β | GCAACTGTTCCTGAACTCAACT | ATCTTTTGGGGTCCGTCAACT |
| Ccl2 | Mcp1 | GCCTGCTGTTCACAGTTGC | CAGGTGAGTGGGGCGTTA |
| Tnfa | Tnfa | CCCTCACACTCAGATCATCTTCT | GCTACGACGTGGGCTACAG |
| Tnfsf10 | Trail | GCTCCTGCAGGCTGTGTC | CCAATTTTGGAGTAATTGTCCTG |
| Hprt1 | Hprt | TCAGTCAACGGGGGACATAA | TGCTTAACCAGGGAAAGCAAA |

## IV.15. Statistics

Results were analyzed using the GraphPad Prism 5.0 statistical software. Data are shown as means ± standard error of the mean (SEM). For comparisons between two groups, parametric Student t test or nonparametric Mann–Whitney test were used. For comparisons between more than two groups, parametric one-way ANOVA test followed by a posteriori Bonferroni test was used.

## V. Conclusions

Our results indicate that CAP is able to reduce CCA progression through induction of DNA damage, which leads to cell cycle arrest and apoptosis of tumor cells, together with potential effects in the immune microenvironment in terms of phenotypic change of TAM. These evidences support the potential usefulness of CAP as a future tool to treat CCA. However, several questions remain to be solved before reaching application in CCA patients. First, effect of CAP on healthy liver cells must be evaluated in preclinical orthotopic models of CCA to assess the level of side damaging effects after a direct CAP treatment to the liver. Moreover, to reach human applicability, the size of the CAP applicating device must be reduced and adapted to the human anatomy and localization of biliary tumors. Therefore, although CAP is a novel promising anticancer "agent", further investigation is needed to include it in the therapeutic arsenal of CCA in the future.

**Author Contributions:** Conceptualization, J.V., F.J., T.D., and L.F.; methodology, J.V., F.J., M.V., H.D., A.A., L.A., E.G.S., and F.M.; validation, J.V., F.J., H.D.; formal analysis, J.V., F.J., H.D., J.A., and T.D.; investigation, J.V., F.J., M.V., H.D., A.A., L.A., and F.M.; resources, J.V., F.J., M.V., H.D., L.A., O.S., T.D., and L.F.; data curation, J.V., M.V., H.D., A.A., L.A., T.D., and L.F.; writing—original draft preparation, J.V., T.D., and L.F; writing—review and editing, J.V., H.D., A.A., L.A., E.G.S., C.H., T.D., and L.F.; visualization, J.V., T.D., and L.F.; supervision, J.V., T.D., and L.F.; project administration, J.V., T.D., and L.F.; funding acquisition, L.A., T.D., and L.F,.

**Funding:** JV and FJ are recipient of the LABEX PLAS@PAR (ANR-11-IDEX-0004-02). This work was funded by the LABEX Plas@par project, and received financial state aid managed by the Agence Nationale de la Recherche, as part of the programme "Investissements d'avenir" (ANR-11-IDEX-0004-02), the program Emergence @ Sorbonne Université 2016, the French Ministry of Solidarity and Health and Inserm, INCA-DGOS-Inserm_12560), the « Région Ile-de-France » (Sesame, Ref. 16016309) and the Platform program of Sorbonne Université. MV is supported by Agence Nationale de la Recherche (ANR-17-CE14-0013-01) and AA by Fondation pour la Recherche Médicale (FRM SPF201809007054).

**Acknowledgments:** We acknowledge Tatiana Ledent and her team from Housing and experimental animal facility (HEAF), Centre de recherche Saint-Antoine (CRSA), Brigitte Solhonne from the histomorphology Platform, UMS 30 Lumic, Centre de recherche Saint-Antoine (CRSA). Annie Munier and Romain Morichon from the Flow cytometry-imaging platform UMS_30 LUMIC, CRSA, Haquima El-Mourabit, Nathalie Ferrand, Jean-Alain Martignoles and Maxime Tenon from CRSA for their help in flow cytometry, and Elisabeth Lasnier for plasma biochemistry dosages (Biochemistry Dept. Saint-Antoine Hospital).

**Conflicts of Interest:** The authors declare no conflict of interest.

# VII. Supplement materials

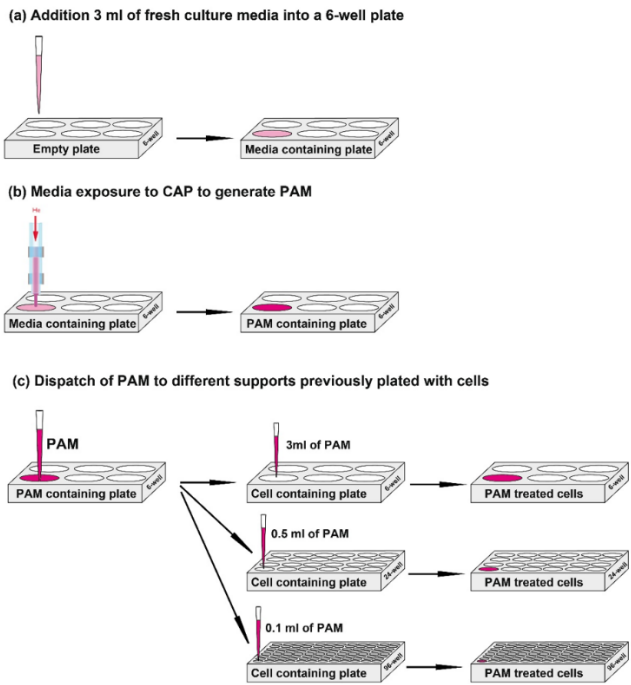

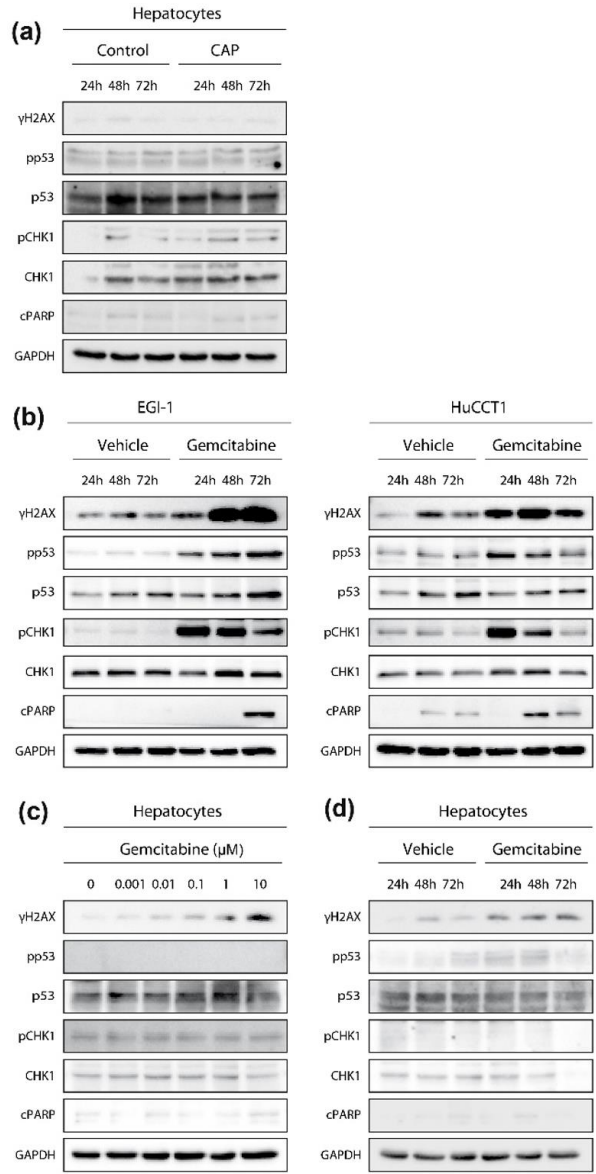

Supplementary Figure 1. Schematic representation of in vitro treatment of CCA cell lines and hepatocytes with PAM. (a) 3 ml of fresh culture media is added into an empty well of a 6-well plate. (b) Fresh culture media is treated with CAP during 1, 3, 5 or 10 minutes to generate PAM. (c) PAM is collected and dispatched in different plastic supports (i.e. 6-well, 24-well or 96-well plates) previously plated with CCA cell lines or hepatocytes. Then plates are developed by different methods at different time points as explained in the manuscript section Materials and Methods.

Supplementary Figure 2. (a) Representative images of western blot analysis of cleaved PARP, phosphorylated and total p53, phosphorylated and total CHK1 and phosphorylated H2AX in hepatocytes after 24, 48h and 72h of exposure to culture medium pretreated with CAP for 3 min (9kV, 30kHz, 14%, gap of 7 mm). (b, c) Representative images of western blot analysis of cleaved PARP, phosphorylated and total p53, phosphorylated and total CHK1 and phosphorylated H2AX in EGI-1 and HuCCT1 after 24, 48h and 72h of exposure to 0.1 and 0.01 µM of gemcitabine, respectively. (d) Representative images of western blot analysis of cleaved PARP, phosphorylated and total p53, phosphorylated and total CHK1 and phosphorylated H2AX in EGI-1 and HuCCT1 after 72h of exposure to increasing doses of gemcitabine. (e) Representative images of western blot analysis of cleaved PARP, phosphorylated and total p53, phosphorylated and total CHK1 and phosphorylated H2AX in EGI-1 and HuCCT1 after 24, 48h and 72h of exposure to 10 µM of gemcitabine.

In the following pages we show the western blots performed with 3 different cultures of EGI-1 and HuCCT1 cholangiocarcinoma (CCA) cells exposed to plasma activated medium (PAM), that were quantified to generate the graphs showed in Figure 7 from the manuscript. The densitometry original data from each of the proteins that were quantified (gammaH2AX, pp53, pCHK1, cPARP and GAPDH) is shown below the corresponding band. We also show the original capture from the PageRuler™ Prestained Protein Ladder (Biorad) taken for each protein and the closer molecular weight marker (PageRuler™ Prestained Protein Ladder, Thermofisher) to each band.

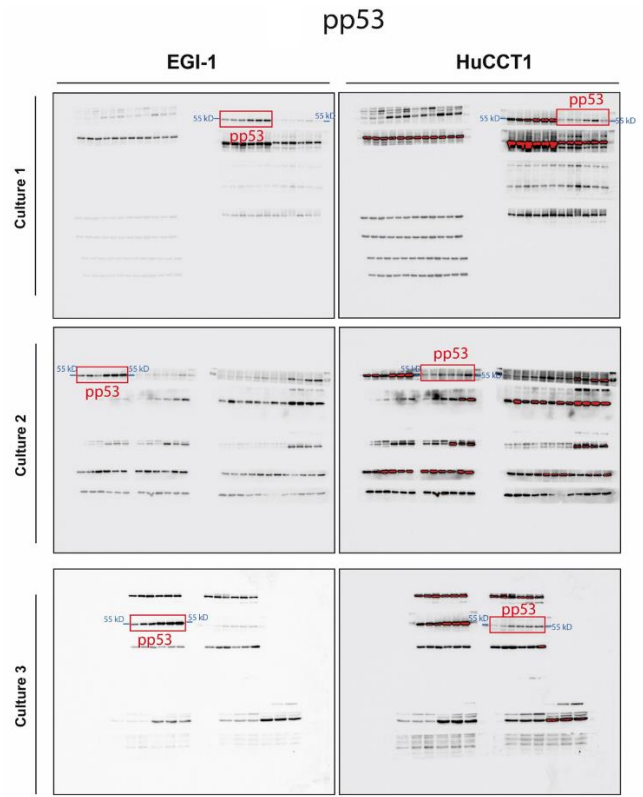

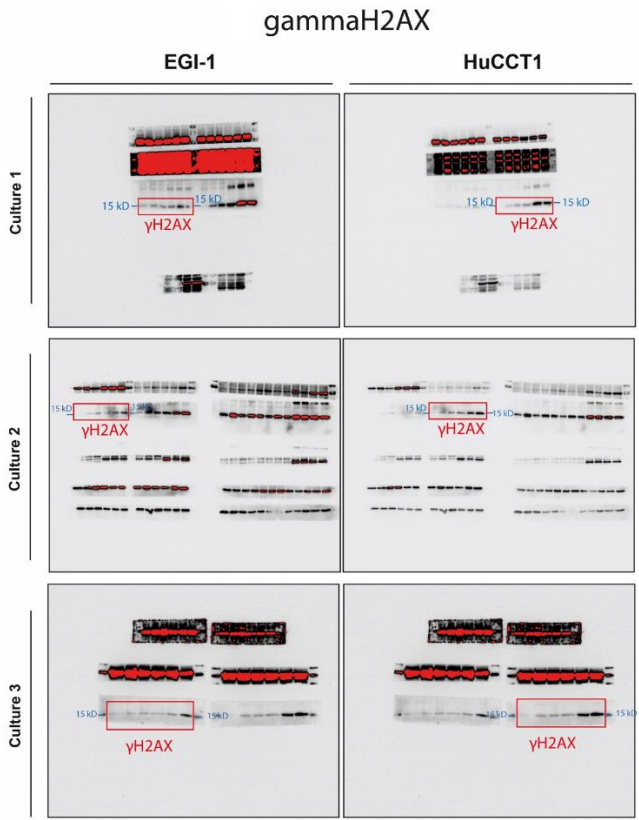

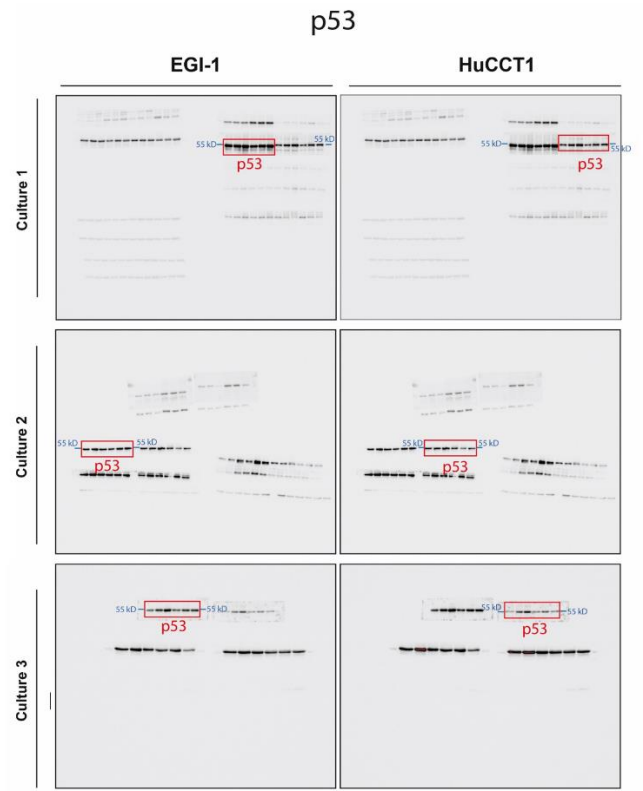

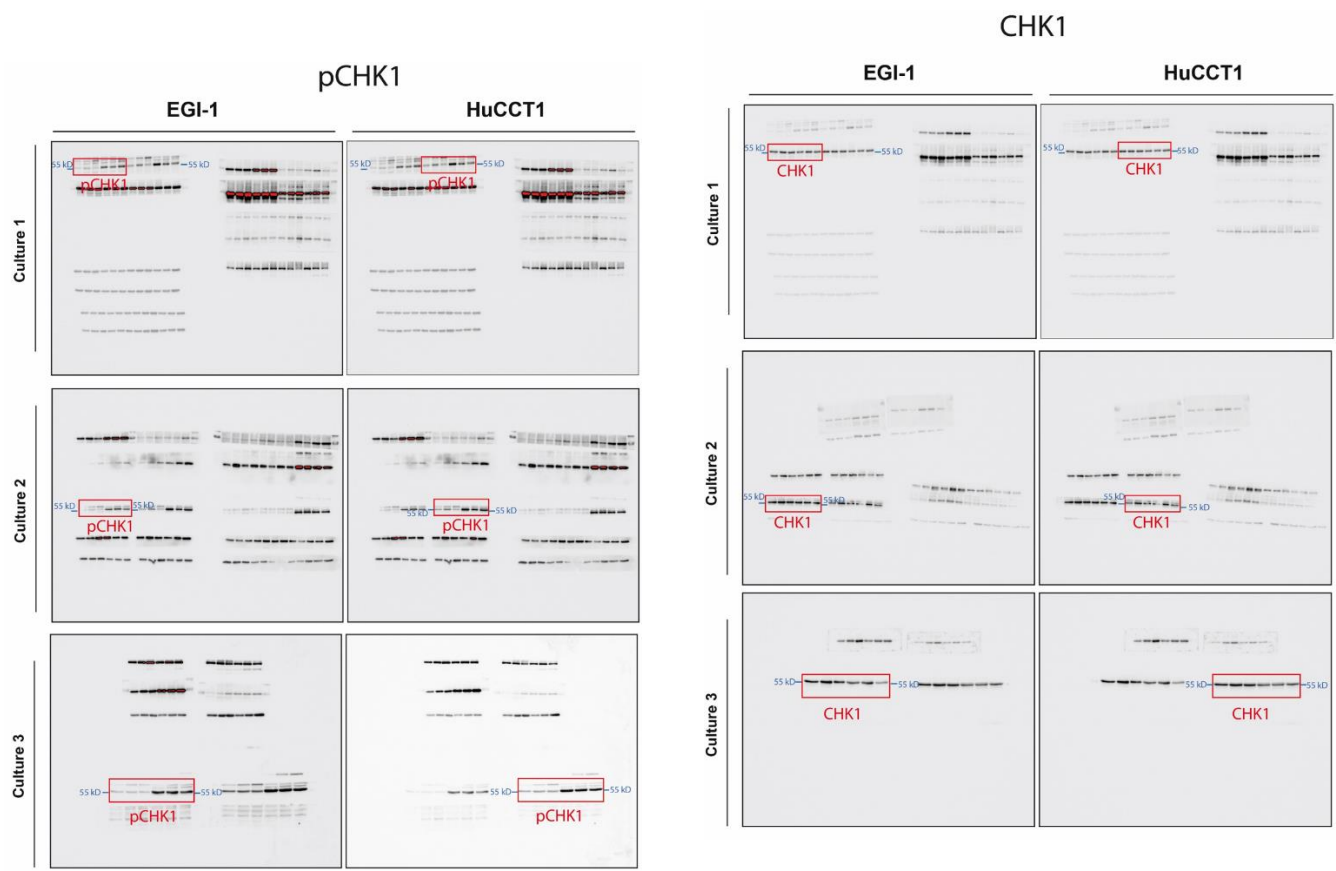

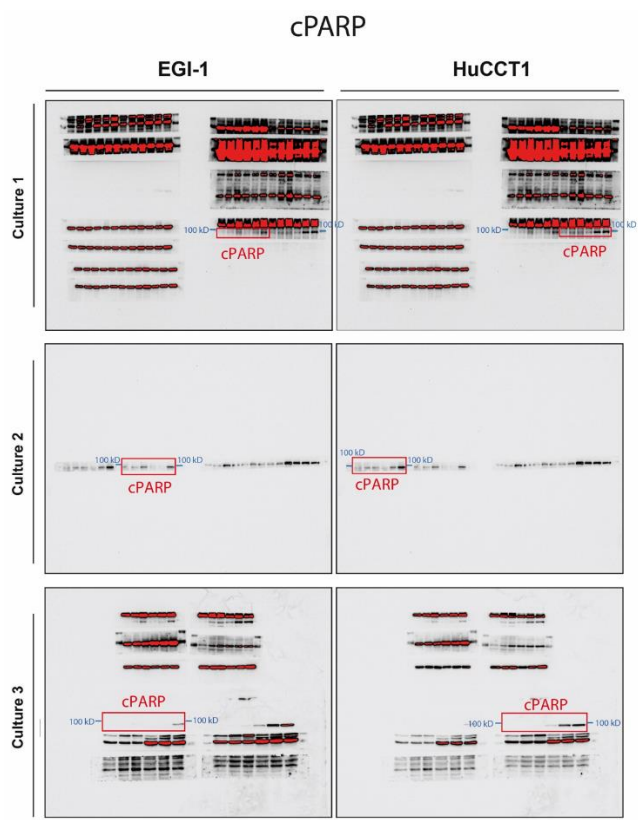
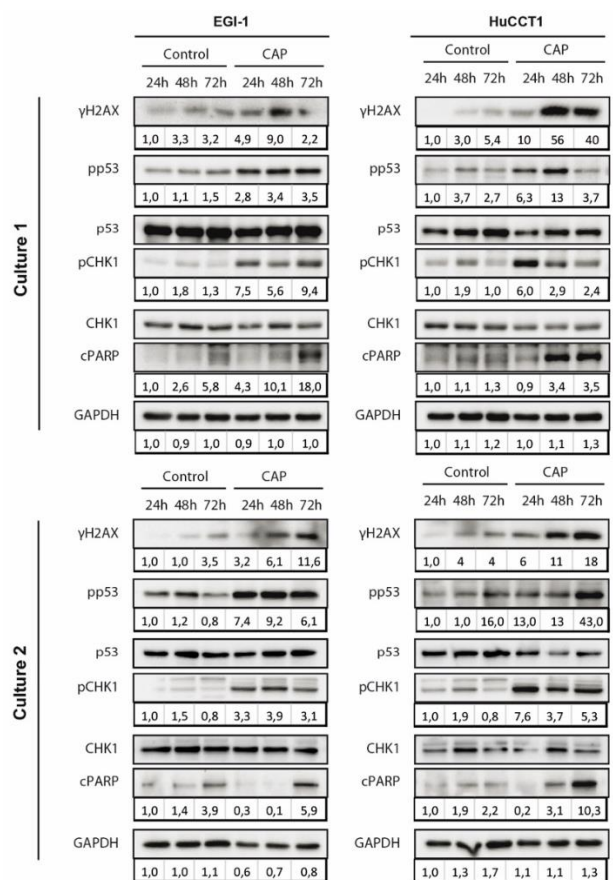
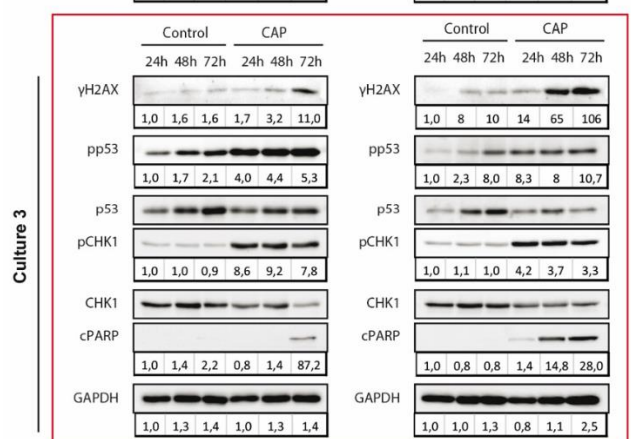